\documentclass[12pt,preprint]{aastex}
\newcommand{\pn}{{\par\noindent}}

\newcommand{\thn}{{\thinspace}}
\newcommand{\scr}{\scriptstyle}
\newcommand{\Msun}{\hbox{$\thn M_{\odot}$}}
\newcommand{\Rsun}{\hbox{$\thn R_{\odot}$}}

\newcommand{\eql}{\thn\thn=\thn\thn}

\newcommand{\w}{{\ \ \ \ \ }}

\newcommand{\x}{\ \ \ }
\newcommand{\by}{\times}
\newcommand{\z}{\ \ }
\newcommand{\on}{{$^1$n}}
\newcommand{\oH}{{$^1$H}}
\newcommand{\tH}{{$^2$H}}
\newcommand{\tHe}{{$^3$He}}
\newcommand{\fHe}{{$^4$He}}
\newcommand{\twC}{{$^{12}$C}}
\newcommand{\thC}{{$^{13}$C}}
\newcommand{\frN}{{$^{14}$N}}
\newcommand{\ffN}{{$^{15}$N}}
\newcommand{\sxO}{{$^{16}$O}}
\newcommand{\svO}{{$^{17}$O}}
\newcommand{\etO}{{$^{18}$O}}
\newcommand{\twNe}{{$^{20}$Ne}}
\newcommand{\toNe}{{$^{21}$Ne}}
\newcommand{\ttNe}{{$^{22}$Ne}}
\newcommand{\ttNa}{{$^{23}$Na}}
\newcommand{\finv}{{F_{\rm inv}}}
\newcommand{\fconv}{{F_{\rm conv}}}
\newcommand{\tn}{{t_{\rm nuclear}}}
\newcommand{\dmm}{{\vert\delta\mu\vert\over\mu}}
\newcommand{\ddmm}{{\vert\delta\mu\vert/\mu}}
\newcommand{\gradr}{ \nabla_{\rm rad}}

\newcommand{\dm}{{\delta\mu}}
\newcommand{\tg}{{\raise 0.3ex\hbox{$\scr  {\thn > \thn }$}}}
\slugcomment{Not to appear in Nonlearned J., 45.}
\shorttitle{Mixing on the Giant Branch}
\shortauthors{Eggleton, Dearborn and Lattanzio}
\begin{document}
\title{Compulsory Deep Mixing of \tHe\ and CNO Isotopes\\ in the Envelopes of low-mass Red Giants}
\author{Peter P. Eggleton, David S. P. Dearborn}
\affil{Lawrence Livermore National Laboratory, 7000 East Ave, Livermore, CA94551, USA}
\email{ppe@igpp.ucllnl.org, dearborn2@llnl.gov}
\and
\author{John C. Lattanzio}
\affil{Monash University, Mathematics Department, Clayton, Victoria, 3168, Australia}
\email{john.lattanzio@sci.monash.edu.au}

\begin{abstract}

Three-dimensional stellar modeling has enabled us to identify a deep-mixing 
mechanism that must operate in all low mass giants.  This mixing 
process is not optional, and is driven by a molecular weight inversion 
created by the \tHe(\tHe,2p)\fHe\ reaction. In this paper we characterize the 
behavior of this mixing, and study its impact on the envelope abundances.  
It not only eliminates the problem of \tHe\  overproduction, reconciling 
stellar and big bang nucleosynthesis with observations, but solves the 
discrepancy between observed and calculated CNO isotope ratios in low 
mass giants, a problem of more than 3 decades' standing.  This mixing 
mechanism, which we call `$\delta\mu$-mixing', operates rapidly (relative to
the nuclear timescale of overall evolution, $\sim 10^8\thn$yrs) once the 
hydrogen burning shell approaches 
the material homogenized by the surface convection zone.  In agreement 
with observations, Pop I stars between 0.8 and 2.0$\Msun$ develop 
\twC/\thC\  ratios of 14.5 $\pm$ 1.5, while Pop II stars process the carbon 
to ratios of 4.0 $\pm$ 0.5.  In stars less than 1.25$\Msun$, this 
mechanism also destroys 90\% to 95\% of the \tHe\  produced on the 
main sequence.
\end{abstract}


\section{Introduction}
In a previous paper (Dearborn et al. 2006; hereinafter Paper I) we used a
fully 3-dimensional code {\tt Djehuty}, developed in the Lawrence Livermore 
National Laboratory, to investigate the onset of the helium flash in a low-mass
red giant. While convective motion due to the helium flash was seen to occur
in much the same  region as predicted by one-dimensional (spherically 
symmetric, hydrostatic)
models, we noticed some minor motion, apparently also of a turbulent convective
character, in an unexpected region: a region above but not far above the 
hydrogen-burning shell, and well below the base of the conventional surface
convection zone. This is visible in Fig.~14 of Paper I. On subsequent
close inspection we found that this additional motion was
due to a very small molecular-weight inversion which developed into
a dynamical instability that we identified as a Rayleigh-Taylor
instability (Eggleton et al. 2006, 2007; hereinafter Papers II, III).
This inversion was due to the burning of \tHe, of which
quite a high concentration is left by the retreating surface convection
zone.

In Papers II and  III
we showed why such an inversion is expected to arise, once the surface 
convection zone, having reached its deepest extent (the `First Dredge-Up', or
FDU), begins to retreat. 
Furthermore, we suggested that this motion should grow in extent so that it 
reaches upwards from the zone where the $\mu$-inversion is maximal (which is
also the zone where \tHe-burning is maximal) to the base of the normal surface 
convection zone. It should lead to the destruction of most 
($\ge 90\%$) of the \tHe\ in the surface layers,
and it should simultaneously allow for some processing of \twC\ to \thC. Thus
at the surface of the star \tHe\ should be progressively depleted, and \thC\ 
progressively enhanced, beyond the values expected from conventional 1D models.
\par In this paper we consider further the effect of our additional mixing,
which we refer to as `$\delta\mu$-mixing'. We would like to emphasise that
this mixing can explain in a natural way the 
observed abundances that have hitherto
been attributed to  mechanisms like rotation and magnetic fields.
We would also like to emphasise the value of 3D modeling. It was only by
using a fully 3D (hydrodynamic) model that this mixing was noticed. It is due 
to a very
slight effect, a $\mu$-inversion that amounts to less than
one part in
$10^4$, and yet it is quite obvious in a 3D simulation. One can in
fact see the inversion in 1D models -- although we are not aware that anyone
has actually commented on it -- but in 1D models mixing only
occurs if the code-developer tells the code to include it.
\par 

Three dimensional 
simulations are expensive of computer time. Our philosophy in using 
{\tt Djehuty} has been that 3D simulations spanning a short period of time
may allow us insight into complex hydrodynamic and hydromagnetic phenomena 
which will then enable us to improve the quality of simplistic 1D models.
We follow this principle here. We introduce into our 1D code a convective 
mixing coefficient that depends on the $\mu$-gradient, but only if the 
$\mu$-gradient is in the sense that it is destabilising. This is in addition
to the mixing coefficient that is due to ordinary convective instability,
which depends (crudely speaking) on the entropy gradient but also only if
this gradient has the sign which is destabilising.
\par In Section 2 we discuss the significance of possible mixing for
abundance measurements on the Giant Branch. In Section 3 we discuss our
new $\delta\mu$-mixing mechanism in more detail. In Section 4 we
describe a 1D code which incorporates a simple model of our 
$\delta\mu$-mixing. 
In Sections 5 and 6 we discuss the sensitivity of this model to some
input parameters that we estimate. In Sections 7 -- 12 we present our results
and conclusions.

\section{The Importance of Mixing on the Giant Branch}

There is a long history of observing the carbon, nitrogen and oxygen (CNO) 
isotope ratios in Red Giants (Lambert \& Dearborn 1972, Day et al. 1974, 
Dearborn et al. 1975, 1976, Tomkin et al. 1975, 1976, Tomkin \& Lambert 
1978, Harris \& Lambert 1984a,b, Gilroy 1989) as a probe of 
stellar interiors and evolution. While on the main sequence, nuclear 
reactions change the abundance distribution in a star's deep envelope. These 
changes are then mixed to the surface in the FDU, when the star becomes 
a giant. Among 
the isotopes that are substantially enhanced by low mass stars are \tHe, 
\thC\ and \svO. Other isotopes like \ffN\  and \etO\  are reduced. 
Distressingly, the observed ratios were nearly always different from what was 
predicted, and sometimes very different (Dearborn et al. 1976, Dearborn 1992).

\par A result of giant-branch abundance changes that is of particular 
consequence 
is the robust result that low-mass stars (below $\sim 1.5\Msun$) are major 
producers of \tHe.  In these stars, the PP chain produces a \tHe-rich peak 
in the envelope of the star.  Giant-branch convection homogenizes this region,
in an episode known as the First Dredge-Up or FDU, 
raising the surface \tHe\ abundance by a factor of $\sim 80$ (see Fig.~1).
As the helium core grows some \tHe\  is destroyed, but only (in classical 1D
models) {\it below} the surface convection zone. The fractional abundance of 
\tHe\ in the convective envelope itself is not diminished, even though the 
mass of the convective envelope is diminished.  
The presence of horizontal branches in the HR 
diagrams of globular clusters can be understood only if substantial amounts 
of envelope mass are lost on the giant branch.  This should result in a 
substantial enhancement of the \tHe\ in the interstellar 
medium, by low-mass stars of both Pop I and Pop II  (Dearborn et al. 1986, 
Steigman et al. 1986, Dearborn et al. 1996).  Work by Hata et al. (1995) 
argued that unless $\ge 90\%$ of this \tHe\  is destroyed prior to ejection, 
the results of stellar nucleosynthesis come into conflict with standard Big 
Bang Nucleosynthesis. To avoid this, there must be some mechanism that operates
well before the helium core flash to destroy the \tHe.   

\par Sweigart \& Mengel (1979), Smith \& Tout (1992) and Wasserburg et al. 
(1995) posited that a deep rotationally-driven circulation mechanism could 
solve 
the problem of low \twC/\thC\  values observed in red giants, and destroy 
the excess \tHe\  at the same time.  Mechanisms have been explored that might 
cause this  
deep mixing, including magnetic-field generation (Hubbard \&
Dearborn 1980) or rotational mixing (Charbonnel 1995, Chaname et
al. 2005). Note in partricular that Palacios et al. (2006) have
shown recently that rotation is unable to produce the required mixing.
The necessity for some deep mixing was also discussed 
by Hogan (1995).

\par To destroy the necessary \tHe, any such mechanism must operate 
efficiently in nearly all low-mass stars.  While rotation is certainly present 
in young stars it is inferred
to decay on the main sequence. The slow rotation 
speeds seen in white dwarfs, and the rotation rate inferred in the Solar core 
by helioseismology, suggest that angular momentum is not conserved in the cores
either. As a consequence, dependence on the ability of rotation to 
destroy the excess \tHe\  in all long-lived stars is unsettling.  Nevertheless,
rotational mixing must surely exist 
in some stars, though of poorly known efficiency.

\par Here, we wish to examine the implications of 
our new mixing mechanism, 
described in Papers II and III, 
that operates efficiently in all low-mass stars.  This new mechanism is not
optional: it inevitably arises on the First Giant Branch when the 
hydrogen-burning shell encroaches 
on the homogenised, formerly convective, zone left behind by the retreating 
convective envelope, and begins to burn \tHe.  This \tHe\ burning occurs just 
outside the normal hydrogen-burning shell, at the base of a radiatively 
stable region about $1\Rsun$ in thickness. The burning causes a molecular 
weight inversion, which drives 
mixing all the way to the usual convection zone.  As we will show, it not 
only routinely destroys $\ge 90\%$ of the \tHe\  produced in the low-mass 
offenders, but reduces the \twC/\thC\ ratios from the range of 20 to 35 
down to the observed values between 5 and 15, depending on metallicity.

\section{The $\delta\mu$-Mixing Mechanism}

As described above, this mixing mechanism is activated when the 
hydrogen-burning shell approaches the homogenized region left behind by the 
retreating surface convection zone. \tHe\ is among the most fragile nuclei 
present, and the reaction 
\begin{equation}
^3{\rm He}\  (\thn ^3{\rm He},\thn 2{\rm p})\ ^4{\rm He} \ ,
\end{equation}
\pn has the unusual characteristic (among fusion reactions in stars) of 
lowering the mean molecular weight $\mu$, creating a localised
$\mu$-inversion. This has already been noted by Ulrich (1972) although he
discussed it only in the context of burning during the pre-main-sequence
contraction phase.\par 

This $\mu$-inversion is only about $0.025 \Rsun$ above the hydrogen-exhausted 
core, and has the lowest molecular weight in the entire star (Fig 2).
We emphasise that the \tHe-burning that we are discussing
produces an inversion
only because of the previous homogenisation of the outer layers by the FDU. 
In the absence of the FDU there would be no
$\mu$-inversion, because the \tHe\ peak would be superimposed on a \oH/\fHe\ 
gradient that gives a strongly stabilising $\mu$-gradient. However when the 
entire outer layers are homogenised by the FDU, and then later encroached on
by the gradual outward (in mass) motion of the burning shell, the background 
gradient has been removed and so the inversion can develop.

Figure 2 shows the development of this $\mu$-inversion as seen in a
1D code, with no instructions for mixing in such regions. The maximum
increase in $\Delta (1/\mu)$ is seen as $\sim 5\times 10^{-4}$. 
In the absence of $\delta\mu$-mixing, the magnitude of the inversion depends
on the \tHe\ abundance. The temperature in the region where \tHe\ is burned is
$\ge 10^7$K, so that the material is fully ionised. The mean molecular weight
is
\begin{equation}
{1\over\mu}=\sum_i\thn{(Z_i+1)X_i\over A_i}\ .
\end{equation}
The \tHe\ cross-section is larger than other rates, resulting in the production
of \fHe\ and \oH\ and reducing the molecular weight. Through reaction (1) 
$\mu$ decreases by an amount that depends on the change in the \tHe\ mass 
fraction. If the mass $X_3$ of \tHe\ in unit mass of gas increases by an amount
$\delta X_3$ (a negative quantity, in the present context) then according to
equation (1) $X_1$ increases by $-2\delta X_3$/6 and $X_4$ increases by
$-4\delta X_3/6$, so that the change in $\mu$ from equation (2) is given by
\begin{equation}
\delta\left({1\over\mu}\right) = -{2\over 1}\thn{2\delta X_3\over 6}+
{3\delta X_3\over 3} - {3\over 4}\thn{4\delta X_3\over 6}
\w{\rm or}\w{\delta\mu\over\mu^2}\eql {\delta X_3\over 6}\ <\ 0\ .
\end{equation}

Figure~1 shows that the initial value of $X_3$ is about $2\times
10^{-3}$ and hence we predict a maximum $\Delta(1/\mu) \sim 3\times
10^{-4}$ in good agreement with the $5\times 10^{-4}$ seen in Figure~2.
It was this model, with the maximum possible value for $\Delta(1/\mu)$
that was mapped into the 3D Djehuty code. The resulting layer was
dynamically unstable  and began to rise at a relatively high speed $\sim
10^2 $m/s.

A simple buoyancy argument was 
used to estimate the rise rate of these clouds to be of order 300~m/s, only 
modestly slower than the speeds expected in the convection region
itself, and in agreement with the speeds seen in the 3D calculation.
As these clouds rise (Figure~3), they are replaced 
with \tHe-rich material, and the process is continued.  If such speeds are 
maintained, the low-molecular-weight material will reach the convection zone 
in a few months. This is to be compared to the hundreds of millions of years 
required for a low-mass star to reach the helium flash after this 
$\delta\mu$-mixing process begins.

\par The disparity of the mixing to evolutionary timescale is such that Paper 
III used an instantaneous mixing approximation and the temperature/density 
structure of the stable region to estimate the amount of \tHe-processing. In 
the time taken by a $1\Msun$ star to evolve up the giant branch, the \tHe\ 
in the envelope was expected to decline by about 3 e-folds or 95\%.  A similar 
estimate for \twC-processing indicated that \twC\  would decrease by about 
8\%. While modest, this is sufficient to reduce the \twC/\thC\  ratio from 25 
to near 15.

\par This analytic calculation suggested that the $\delta\mu$-mixing might 
solve more than the conflict between the Big Bang nucleosynthesis and stellar 
evolution as to which produced the \tHe.  It has the potential to solve many 
of the CNO isotope anomalies. Weiss \& Charbonnel (2004) have identified the 
point where CNO composition differences (from expected values) begin to occur 
as `where the hydrogen-burning shell encounters the deepest point to which the 
convective envelope ever reached', and this is just where the 
$\delta\mu$-mixing mechanism becomes operational.

In reality, however, the estimate in Paper II and the mixing seen
in the 3D calculation is likely a substantial  overestimate for the
following reason. Once the $^3$He starts to burn it will drive mixing of
the low-$\mu$ material and hence the actual $\mu$-inversion will be less
than the maximum case discussed above. The actual value of $\delta \mu$
will depend on the mixing speed, which we try to estimate below.

To examine this phenomenon
quantitatively, we have incorporated a 
$\dm$-mixing model into a 1D stellar evolution code.  The code tracks 
16 isotopes, 
including all stable isotopes of the CNO tri-cycle: \oH, \tH, \tHe, \fHe, \twC,
\thC, \frN, \ffN, \sxO, \svO, \etO, \twNe, \toNe, \ttNe, \ttNa, \on.  They are 
coupled through reaction rates taken from Caughlan \& Fowler (1988). Following 
their recommendation, in reactions like \svO(p,$\alpha)$\frN\ and 
\svO(p,$\gamma$)$^{18}$F that have an uncertain factor (0 to 1) on certain 
states, the factor was chosen to be 0.1. As discussed by Dearborn (1992), 
these factors are significant for the expected oxygen isotope ratios. In the 
sections below, this model will be tested and used to explore a range of 
masses and metallicities. Uncertainties in the reaction rates, and their effect
on stellar evolution, have been discussed recently by Herwig et al (2006).

We sum up the overall mixing that we expect as follows:
\pn (a) A stably stratified layer reaches temperatures where $^3$He
begins to fuse with itself, decreasing the molecular weight of the layer
{\it in situ}. Rather than develop as a dynamical instability
(identified as a Rayleigh-Taylor instability in Paper II) the mixing is
more like a thermohaline process, with the buoyancy determined by the
competition between the diffusion of heat from the layer on the one
hand, and the mixing (``diffusion'') of the $^3$He fuel on the other
hand. To the extent that the mixing is a diffusive process (and it is
not) one could call this a ``double diffusive'' mixing, much like the
thermohaline mixing seen in salty water (Stern 1960; Veronis 1965).
In the case of salty water, the mixing is determined by the competition 
between the diffusion of salt on the one hand and heat on the other.
In our case the competition is between the mixing of the material
and the heat diffusion.

\pn (b) In addition, normal convection drives mixing throughout the classical
surface convection zone. This mixing is much  more rapid than the
$\delta \mu$-mixing discussed above. Classical convective mixing is likely to be reasonably well-modeled
by a diffusion process, with a diffusion coefficient $D\sim wl$ (Eggleton
1973, 1983) where $w$ is the mean turbulent speed of convection and $l$ is a
`mixing-length' comparable to the pressure scale height.

\pn (c) Process (a), however, is clearly not a process that should be
well-modeled by diffusion. It might be more reasonably considered advective
rather than convective, and perhaps better characterised by a speed rather
than a quantity like $D$ of dimensions speed $\by$ length.  This
is because we expect  finger-like structures to form, which will slowly
mix and will not really resemble a homogeneous mixing process, as 
results from the diffusion equation. In the next
two Sections we consider some estimates of the speed, and an estimate for
an artificial $D$ that will lead in practice to the sorts of speeds that
we  estimate.

A number of papers have discussed mixing in the presence of a
molecular weight inversion. By analogy with the situation seen in salt
water, where surface evaporation creates a layer of 
hot salty water atop fresher
cooler water, such situations are usually called ``thermohaline'' mixing.
A theory for this mixing has been developed by Ulrich (1972) as well
as Kippenhahn et al (1980) - see also the discussion in Kippenhahn (1974).
The key result from these studies is that the mixing time-scale
varies as the square of the size, $d$, of the unstable region. 
In the limit where $d$ approaches the size of the
star, the mixing timescale becomes the Kelvin-Helmholtz timescale.
A rather unfortunate feature of the theoretical work is that 
estimates of the
diffusion co-efficients vary by almost two orders of magnitude between
authors, reflecting uncertainties in the geometry of the motion. Rather
than place the emphasis on the existing attempts at modelling
thermohaline mixing in stars, we prefer to take a phenomenological
approach and estimate mixing speeds from first principles.

\section{The Speed of $\delta\mu$-Mixing}

Our {\tt Djehuty} calculation began from an artificial situation, in
that the $\mu$-inversion had been allowed to grow to its maximum size
already in the previous 1D calculation. We therefore saw rapid
motion setting in rather quickly. In practice, the motion should have started
as soon as the inversion began, and the motion set up by that, though
slower, would have prevented the inversion building up to the size that
we see in Figure~2. We attempt to make a somewhat more realistic estimate of the 
slower motion that we would expect to be set up in a roughly steady state.
(Note that the 3D simulation did show rising blobs of material, but did not have
the resolution to provide us with any guidance to the aspect ratio of the
expected ``fingers''. We may investigate this in the future, although 
{\tt Djehuty} is unlikely to provide much guidance on this question
because of the expected long timescale being well 
beyond that of a hydrodynamic code like {\tt Djehuty}.

\par In a real stellar environment the
buoyancy will begin when only a fraction of the \tHe\ is processed, and it will
move only a short distance before coming to hydrostatic equilibrium. When the
bubble reaches this new equilibrium, its temperature will be lower than its
surroundings by an amount
\begin{equation}
{\delta T\over T}\eql{\delta \mu\over \mu}\ .
\end{equation}
Heat will diffuse in, and the bubble will continue to rise on a thermal
timescale. As a crude overestimate we can take the mixing time to 
be of the same 
order as the thermal timescale $\tau$, defined as the time it takes for the
thermal energy in the radiative 
layer ($3.65\times 10^{46}$ ergs)  to be replaced given the current 
luminosity ($33.2L_\odot$), giving $\tau \sim 5000\thn$ yrs. This 
gives an estimate (probably a considerable underestimate) of the mixing speed:
\begin{equation}
 v\ \ge\ {\Delta r\over \tau}\sim 0.2\thn{\rm cm/s}\ .
\end{equation}
where $\Delta r$ is the thickness of the zone ($\sim 1\Rsun$) between
the $\mu$-minimum and the conventional surface convection zone.
While this mixing is not a diffusion process, the time for material to rise 
through the stable region is orders of magnitude shorter than the time for 
the hydrogen shell to burn through that material. As at result, this region 
maintains homogenization with the convective envelope.  This allows a simple 
diffusion approach, and in the next section, we develop a diffusion 
approximation based on $\dm$.  We will 
show that increases in this diffusion 
coefficient by a factor of 100, or decreases by a factor of 30, have only a
modest effect ($\sim 30$\%) on the determined abundances. 

We attempt to make a somewhat sharper estimate of the speed with which
the bubble will rise. Consider a bubble of radius $l$ containing lower-$\mu$
material. To obtain an upper limit to the velocity we will assume
that the moving bubble is optically thin.
The rate with which heat energy enters the bubble can be estimated
as 
\begin{equation}
F \sim 4\pi l^2\thn.\thn ac T^3 \delta T \sim 4\pi l^2ac T^4
{\vert\delta\mu\vert\over\mu}\ .
\end{equation}
As flux enters the low-temperature region, the temperature increases at a rate
that depends on the volume:
\begin{equation}
F\sim {4\over 3}\thn\pi\thn l^3\thn{3\over 2} {\rho N_0k\over \mu}
\thn{dT\over dt}\ ,
\end{equation}
with $N_0$ being Avogadro's number and $k$ Boltzmann's constant, and so
\begin{equation}
{dT\over dt}\sim{2ac T^4\over \rho N_0 kl}\thn\vert\delta\mu\vert\ .
\end{equation}
The temperature gradient that must be overcome is
\begin{equation}
{dT\over dr}\eql{T\over P}\thn\nabla_{\rm rad}\thn g\rho\ .
\end{equation}
This is accomplished as energy flows into the bubble resulting in a rise
velocity of
\begin{equation}
v\ \sim\ {dT/dt\over dT/dr}\ \sim\ {2ac T^4\over gl\rho \gradr}\thn\dmm
\ .
\end{equation}

\par The process of thermally-driven buoyancy described by equation (10)
should leave the temperature gradient radiative, but we expect this to lead 
to elongated vertical structures resembling
`salt-fingers' (Wilson \& Mayle 1988, Dalhed et al. 1999). As energy  diffuses
into the outer portion of the bubbles, the material rises, exposing the inner 
material.  These slender structures have more surface area per unit volume 
than the bubble model used here, allowing energy to diffuse more rapidly into 
the low-$\mu$ material, and this results in higher rise velocities than our 
estimate here.

The factor $\vert\delta\mu\vert/\mu$ is set at the deepest level of the 
mixing region where the heat from the H-burning shell drives the \tHe
\ reaction. This gives
\begin{equation}
\dmm\eql-{\mu\delta X_3\over 6}\eql{l\mu\over 6v}\thn{dX_3\over dt}\eql {1\over 2}
N_0\rho\left({X_3\over 3}\right)^2R_{\rm nuc}\thn {l\mu\over 6v}\ ,
\end{equation}
where $R_{\rm nuc}$ is the thermal average of reaction cross-section
times speed. Then combining (15) and (16),
\begin{equation}
v^2\eql{acT^4N_0 R_{\rm nuc}\mu\over 6g\gradr}\left({X_3\over 3}\right)^2
\ .
\end{equation}

Evaluating these equations in the region where $^3$He is burned for a 1$\Msun$ 
Pop I model, gives an initial speed in the bubble formation region of 
$\sim\thn 2$ cm/s.  This 
is an order of magnitude above the first crude (under-)estimate, but well within the 
range of diffusion coefficients tested in the next section.  The speed in the 
warm processing region is limited by the conversion rate of $^3$He and it is 
here that $\dm$  is established, following which 
it can be held constant and equation (10) used
to calculate the speed. This speed will be typically somewhat larger
than the initial speed.

\par Averaged over the classically stable region, equation (10) in Pop I 
models show speeds of $\sim 1 - 2\thn$m/s. These high speeds are irrelevant to 
the processing region and only serve to speed homogenization with the convection 
zone. Pop II models
develop lower $\ddmm$ values, and average velocities nearer $0.5\thn$m/s. This
estimate was repeated for several core masses as the model evolved up the FGB,
resulting in mixing times of between 10 and 50 years. Models of different mass 
were also examined, with the result that lower-mass models mix somewhat more 
slowly and higher-mass models somewhat faster. Still, for Pop I models the
mixing timescale was under 100 years (Figure~4). We note that the
difference in mixing speed at the bottom of the envelope, and the exact 
location of the peak in the $\mu$ inversion, will determine the final 
carbon isotope ratio. As we discuss below, these details require
further investigation and are beyond the scope of the current paper.

\par Extreme Pop II models ($Z=0.0001$) were also examined. The initial mixing 
in these models starts at higher temperatures, resulting in greater values of 
$\ddmm$. The mixing timescales start near 10 years for all the masses 
examined. However, as these models evolved the timescale rapidly increased to
$\sim 100\thn$years. This is probably a result of the rapid destruction of
\tHe. A mixing time of 100 years is about $0.01\%$ of the time to burn through 
the classically stable region. We can argue that the classically stable region
will be homogenised with the classically convective surface zone on a short
timescale.

\par We do not feel that simplistic analyses such as equations (6) to (12)
are a very reliable guide to the velocities to be expected, and in particular
we do not feel that a roughly mixing-length-like theory is appropriate for
the rate of travel of salt-finger-like mixing. In the process described after
equation (10), based on 3D simulations in the papers quoted, the bubble
does not in fact wait for all the heat to diffuse in to the center before
rising starts. That appears to be a key factor in the finger-like character
of the structures. Because of that we have used an optically thin approximation
in equation (6) rather than an optically thick approximation. But several
other steps in this kind of back-of-the-envelope estimation are quite
problematic. We reiterate that although theories for
thermohaline mixing exist, they are uncertain by typically two orders
of magnitude, as reflected in the preferred diffusion co-efficients used
by Ulrich (1972) and Kippenhahn et al (1980).

\par We prefer (next
Section) to use an alternative simplistic procedure, which is to choose a
$\dm$-mixing diffusion coefficient such that it will give speeds of advance
comparable to the kinds of speeds we have estimated above. We are 
fortunate that it does not matter, to within several orders
of magnitude, just how rapid is the actual motion. 
As long as 
the motion is rapid compared to the very slow overall nuclear timescale of
the star then  the results will be essentially the same, and independent
of the mixing speed.
Whether the speed of rising elements is 2 m/s or 0.2 cm/s, this
is still faster than the speed with which material would have to travel to
mix on a nuclear timescale, $\sim 10^{-3}$cm/s. Indeed, this is
confirmed below where we show that varying the mixing speed by more than
three orders of magnitude provides quantitative changes of order $\sim$
30\% in the determined surface abundances.

\section{The $\delta\mu$-Mixing Model}

In 1D codes, mixing always requires some physical model to approximate the 
process. Our 1D code treats normal convection as a diffusion process and 
solves a second-order equation for each isotope: 
\def\mr{{(4\pi r^2\rho)^2}}
\begin{equation}
\left({\partial X\over\partial t}\right)_k={\partial\over\partial m}\thn
D\thn\mr\thn{\partial X\over\partial m}+R+{\partial X\over\partial m}\thn
\left({\partial m\over\partial t}\right)_k\ .
\end{equation}
The first term in the equation for the rate of change of the isotope is 
for the convective diffusion, the next term ($R$) incorporates the nuclear 
reaction rates, and there is a final term to deal with the mesh motion,
because the mesh is non-Lagrangian; $k$ is the meshpoint number.

\par In a standard convective region the diffusion coefficient $D$ that 
we use takes a form that scales quadratically with the temperature-gradient 
excess over the adiabatic value, and inversely with the nuclear timescale:  
\begin{equation}
 D={\fconv r^2\over\tn}\left[\max(0,\nabla_r-\nabla_a)\right]^2\ ,
\end{equation}
where $\nabla_r$ and $\nabla_a$ are the usual radiative and adiabatic
temperature gradients from the mixing-length theory of convection. The 
timescale $\tn$ is an estimate of the nuclear evolution timescale: $\tn=
0.1EX_0M/L$, where $E$ is the nuclear energy available from hydrogen burning,
$X$ is the abundance of hydrogen in the outer layers ($\sim 0.7$) and $M$
is the mass of the star: $\tn \sim 10^{10}\thn$yrs for the ZAMS sun, and is
$\sim 10^8\thn$ yrs in the middle of the FGB.
The dimensionless and largely arbitrary factor $\fconv$\  is simply chosen 
to be a large number such that the composition in a convective region 
homogenizes in a time much shorter than the nuclear time scale ($\tn$). 
We find that a value of $10^6$ works well. A larger value would be more
physical, but would lead to numerical difficulty since the stepwise
difference in composition becomes so small as to be susceptible to
rounding error. There is little difference in practice between a convective
region being homogenised in $10^{-6}\tn$ and $10^{-10}\tn$. 
\par To model our $\delta\mu$-mixing process we have created a diffusion 
coefficient wherever there is an inversion. The form that we have used
is simple, and in analogy with the previous equation:
\begin{eqnarray}
D&=&{\finv r^2\over\tn}\thn(\mu - \mu_{\rm min})\ \ \ (k\ge k_{\rm min})\\
& =&0\ \ \ (k\le k_{\rm min})\ ,
\end{eqnarray}
where (a) $\mu_{\rm min}$ is the smallest value of $\mu$ in the current model,
(b) $\mu_{\rm min}$ occurs at meshpoint number $k_{\rm min}$, and (c) $k$,
the meshpoint number, is counted outwards from the center. 
Thus we introduce mixing wherever there is an inversion  in $\mu$,
in the same form as usually used for mixing, but with a factor $F_{inv}$ to be determined.  
In a region where 
both of equations (14) and (15) give $D\tg 0$, we only use (14).
Again, the factor $\finv$, if large, simply assures homogeneity in a time much 
shorter than the nuclear time scale.  It remains for us to
determine a value
of $\finv$ that is appropriate for $\delta \mu$-mixing.

\par To estimate the speed that corresponds 
to a chosen value of $\finv$, we performed a numerical test in which a step 
function was installed in an element that was not being used in the 
nucleosynthesis network.  The position of the step was located just above the 
point where the $\mu$-inversion would form.  Below this point, the mass 
fraction was set to 10$^{-7}$, and above this point dropped to $10^{-10}$.  
Once the inversion develops, 
mixing begins, and a stable gradient is formed. The rate at which the material 
below the step is transported to the surface was then monitored to obtain an 
effective speed (with the usual convective diffusion coefficient turned off to 
avoid confusion). This is illustrated in Figure~5, for $\finv=10^2$. Some
material is seen to have travelled $\sim 6\times 10^{11}$cm in 80,000 yrs,
ie. with a speed of $\sim 0.3\thn$cm/s. This is about as slow as our slowest 
estimate above, and hence is chosen as our standard
value. Increasing 
$\finv$\  by 10 times results in a speed that approached 1.5~cm/s.  Increasing 
this arbitrary value by another factor of 10 (100 times the standard 
value) increases the speed to 6~cm/s.  These speeds 
are sufficient to mix the stable region with the outer 
convective region on a time that is short in comparison to the evolution.  
These speeds are near the startup velocity found where the reduced 
$\vert\delta\mu\vert/\mu$ is established, where the 
nucleosynthesis is done, and where the composition changes are most 
sensitive to mixing speed.

\par Figure~6 shows the diffusion coefficient that we used for $\dm$-mixing,
along with the diffusion coefficient used for the ordinary convective 
envelope. Beyond the point where the switchover occurs, we plot 
(and use) only our value for normal convective mixing. We note that
$\delta \mu$-mixing produces diffusion co-efficients that are 
up to two orders of
magnitude larger than found for rotating models (Palacios et al 2006).

\par To illustrate the effect  of this 
$\delta\mu$-mixing on a star's evolution, 
a $1\Msun$ ($Z=0.02$) model was evolved from a 
pre-main-sequence configuration to the helium core flash.  The initial 
\twC/\thC\  ratio was chosen to be 90, and as the model reached the giant 
branch ($M_{\rm core}\sim 0.2\Msun$) the first dredge-up reduced the ratio to 
29.5.  When the new mixing mechanism began, the \twC/\thC\  ratio declined 
rapidly to near 15 (Figure~7). At this point, 
the reduced \tHe\  abundance slowed the mixing to the extent that between a 
core mass of 0.3 and $0.45\Msun$ the ratio declined only to 14.3.  Nearly 93\% 
of the initial \tHe\   was destroyed by the tip of the Giant Branch.

\par When the $\delta\mu$-mixing is active, the striking molecular weight 
inversion (red in Figure~8) does not develop.  Instead, a much more gradual 
profile is developed with a modest cusp in the region where the \tHe\  burning 
occurs. We find $\Delta(\mu/\mu) \sim 10^{-5}$ but the details will
depend on the assumed mixing speed as well as the temperature of the
approaching shell.

\section{Sensitivity of the \twC/\thC\ ratio and \tHe\  destruction to 
$\finv$}

It is important to realise that  our preferred value of $\finv$ was
chosen to match our estimates of the mixing velocity. Nevertheless, we
now wish to test the sensitivity of the \tHe\  destruction 
and the final \twC/\thC\  ratio to the 
value of  $\finv$.  A series of runs were made on a $1\Msun$ $Z=0.02$ model.
The value of $\finv$\ was varied over a factor of 10,000 (from 0.01 of the 
standard value to 100 times the standard value).  To test the sensitivity to 
mesh resolution, this test was done for models with both 300 zones 
and 750 zones. 

\par Table~1 shows the effect of varying $\finv$ by factors of up
to $10^4$. For values of
$\finv \lesssim 10$ the mixing is too slow 
and the \tHe\  burning is incomplete. For rates near the standard value 
chosen, destruction is near 90\%, and \twC/\thC\  ratios are a minimum. 
Larger diffusion coefficients only modestly reduce the \tHe\  abundance and 
give a little less reduction in \twC/\thC. The differences are not
significant. Note that the preferred value of $\finv = 100$ was chosen 
becasue it
gives a mixing speed close to the minimum that we estimated; hence $\finv
= 10$ is an order of magntude slower than our estimate of the minimum.
The table also shows a modest sensitivity to mesh resolution. This
information is repeated more graphically in Figure~9.

Next we examined a range of masses (for $Z=0.02$, and 300 zones) 
comparing diffusion coefficients that differ by a factor of 100. 
For the higher diffusion coefficients 
the \tHe\ destruction is almost identical except for 
the higher mass where less \tHe\ is produced in the first place.  
There are modest differences in \twC/\thC\  ratios that at present 
should be considered as uncertainty in the modeling of the 
mechanism.

Table 2 shows that the behavior seen in the $1\Msun$ model holds over the 
range of interesting masses. Between 0.8 and 2.0$\Msun$, higher diffusion 
coefficients result in small changes in the \tHe\  destruction, and very 
modest differences in \twC/\thC\  ratios. 
The differences in \twC/\thC\ ratios caused by varying mesh and mixing 
coefficient ($\pm$2 for Pop I models) should be considered uncertainty 
in the modeling of the mechanism (at present).  When the ratio drops 
to near the equilibrium value of 3.5, as we will see in Pop II models, 
these factors have much less effect on the \twC/\thC\  ratios. 

\section{\twC/\thC\  and $\delta\mu$-Mixing}

The following section examines the \twC/\thC\  ratios, and the 
helium production, for a range of low-mass stars with Pop I and 
Pop II metallicities.  We use the standard diffusion coefficient 
($\finv = 100$) developed above for the $\delta\mu$-mixing.  Consistent 
with Anders \& Grevesse (1989), the initial \tHe\ mass fraction 
was taken to be $2\by 10^{-5}$. While it is appropriate to use a higher 
value to account for the conversion of D to \tHe\ on the pre-main 
sequence, this difference is minor when compared to the main 
sequence production of low-mass stars.

\par Table~3 shows the \twC/\thC\  ratios for models ranging from 
0.8 to $2.0\Msun$, and for metallicities from solar to 1/50th solar.  
In the absence of an additional mixing process, the final (tip of Giant 
Branch) value of the carbon isotope ratio depends on mass. As the mass 
rises from 0.8 to $2.0\Msun$, the expected \twC/\thC\  ratio drops from 
near 35 to near 20, with very little dependence on $Z$.  This mass 
dependence is seen in the FDU columns of Table~3, showing the 
\twC/\thC\  values after FDU and before $\delta\mu$-mixing begins.  Once the 
mixing begins, the \twC/\thC\  ratio rapidly drops to a lower 
value, and the final range of ratios (`Final' in Table~3) show a considerably 
reduced range. For solar metallicities, stars in this mass range all 
show \twC/\thC\ $\approx$ 14.5 $\pm$ 1.5. Similarly, for $Z$ = 1/50th 
solar, \twC/\thC\  $\approx$ 4.0 $\pm$ 0.5. The $2.0\Msun$ models are 
not included in these averages, as the $\delta\mu$-mixing begins just 
prior to helium core flash in the Pop I model, and has not begun 
in the $Z$ = 0.0004 model.

That the final \twC/\thC\  values converge for a broad range 
of masses is an interesting result, and is shown graphically in Figure~10. 
Before $\delta\mu$-mixing begins, the carbon isotope ratios show the usual 
mass dependent range, but not afterwards. To illustrate $Z$ dependence in 
more detail, a star of mass $0.9\Msun$ was evolved with various values of 
$Z$ between solar and 1/200th solar (Figure~11). The \twC/\thC\  ratio is 
seen to 
vary smoothly from 14.8 to 3.5. Additionally there are big changes in 
\frN/\ffN, and small changes in O isotope ratios (mostly due to \etO\  
and \svO). Table~4 provides the same information in tabular form.

In the low-mass stars of interest here, the PP chain dominates evolution on 
the main sequence, but the hydrogen-burning shell on the giant branch 
operates on the CNO cycle. Stars with a lower metallicity Z will also have
fewer CNO nuclei, so that the shell must burn at a 
somewhat higher temperature for the same energy production rate.  
Additionally, the penetration of the surface convection region is not as deep 
at low $Z$, and the inversion is not initiated until the core grows to a 
larger mass. These effects combine to result in a higher temperature in the 
place where the molecular weight inversion develops. The position
of this inversion is the result of competition between the advance of
the hydrogen-burning shell (whose temperature is dependent on the
star's metallicity) and the speed of mixing. As $Z$ decreases from 
0.02 to 0.0001, the core mass at which the inversion occurs increases 
(from 0.233 to 0.358 for models of mass $0.9\Msun$). The luminosity, $\log L$, 
at the start of mixing increases from 1.4 to 2.4, and the temperature at the 
base of the mixing increases from 16.5 million K to 24.0 million. As a result, 
the mixing begins at a metallicity-dependent temperature, and the \twC/\thC\  
ratio achieves different values before the reduced \tHe\ abundance slows 
the process (Figure~12).

It is again important to note that the composition dependence of
our results was in no way specified by us, but is a direct result
of the composition  dependence of the temperature in the burning regions.
All we have done is determine a mixing velocity (with no explicit
composition dependence). The result is exactly as observed:~the carbon
isotope ratios fall further for lower metallicities.

As a final note, these models were all run with a mixing length that was 
1.8 pressure scale heights (to fit the solar radius). Changing the mixing 
length $\pm$ 0.2 resulted in changes in the \twC/\thC\  ratio that were 
$<$0.4.  

\section{The Metallicity Effect and Observations}

Gilroy \& Brown (1991) measured the carbon isotope ratios in the 
stars of M67 (near 1.2$\Msun$).  They report that `the subgiants 
seem to have undergone little or no mixing', and that the lower 
giant branch stars exhibit `normal first dredge-up mixing ratios'.  
However they find 8 upper giant branch and clump giants with 
\twC/\thC\  ratios between 11 and 15.  This is in excellent 
agreement with the value near 13.5 that results from $\delta\mu$-mixing 
(Table~3).  At the lower metal enrichments, Pavlenko et al. 
(2003) observed giants in the globular clusters M3, M5, and M13 
([M/H] = $-1.3, -1.4$, and $-1.6$ or $Z\le 0.001$), finding 
\twC/\thC\  between 3 and 5.  Again this is in excellent accord 
with our models (Table~3).  They also observed M71 ([M/H] = $-0.71$ 
or $Z\approx 0.004$) finding less processing.  Here the \twC/\thC\  
ratios show values near 7 (5 to 9). The $0.9\Msun$ model that we 
used to illustrate the metallicity effect gives an expectation for 
\twC/\thC\  ratios near 8.

\section{\tHe\ and $\delta\mu$-Mixing}

As stated earlier, in the calculations done here the initial mass fraction 
of \tHe\ was taken to be $2\by10^{-5}$.  In Table~5, a $0.9\Msun$ model 
is evaluated for a range of metallicities to find the peak enhancement in the 
mass of \tHe, M(\thn\tHe, peak)/M(\thn\tHe, init).  For solar composition, 
the enhancement reaches a factor of 61.  With shorter main sequence 
lifetimes, Pop II models of this mass reach only about 38. Increasing the 
main sequence value of \tHe\ by an order of magnitude, to $2\by 10^{-4}$, by 
assuming all of the deuterium is burned to \tHe\ on the pre-main sequence 
makes surprisingly little difference.
For Pop I abundances, starting with an enhanced initial \tHe\  abundance 
associated with converting \tH\ to \tHe\ results in a peak enhancement of 64 
times the ISM \tHe\  used to form the star ($2\by 10^{-5}$) instead of 61.  
For an extreme Pop II metallicity, the peak enhancement is 41 instead of 38.

These very large peak enhancements were problematic in reconciling stellar 
nucleosynthesis and Big Bang nucleosynthesis with observed abundances.   
With $\delta\mu$-mixing the final enhancement for Pop I abundances is 
reduced to 3.1 (3.4 when \tH\ conversion is included), and for Pop II 
models the final value is 0.52 (0.53 with \tH\ conversion). 

Table~6 shows the same behavior for stars in the mass range of 0.8 
to $2\Msun$. Table~6 shows the ratio of the peak Pre-mix mass of \tHe\ 
to the original mass of \tHe\ in the star, as well as the post-mixing ratio

As found in earlier papers, the greatest potential \tHe\ enhancement 
occurs for stars of $1\Msun$ and below.  In these models, it is 
the large \tHe\ enhancement that enables the $\delta\mu$-mixing 
to destroy 90 to 95\% of the potential \tHe\ contribution to the 
interstellar medium.  As a final note, at and above  $\sim 2\Msun$ the 
$\delta\mu$-mixing operates to (near) completion only for 
the Pop I model.  In the $Z$ = 0.001 calculation, the mechanism begins 
just prior to the helium flash and is incomplete.  In the $Z$ = 0.0004 
model, the helium core flash precedes any significant 
$\delta\mu$-mixing.

\section{\sxO\ and \ttNa\ with $\delta\mu$-Mixing}

Some observational evidence has suggested that \sxO\ and \ttNa\ may 
change on the upper giant branch.  In the $1.5\Msun$ models 
run with $Z$ = 0.02 and 0.0001, there was substantial \ttNa\  
enhancement: $\delta$(\ttNa) $\approx$ 19\% and 68\%.  However, 
this was just the usual enhancement expected from main 
sequence processing followed by giant branch mixing, and 
is not significantly effected by $\delta\mu$-mixing.  
When we examined the $0.9\Msun$ model the greatest 
change was seen for the extreme Pop II abundance (1/200th solar), 
where the \ttNa\ abundance climbs by 3\%.  Only about half of 
this change is due to $\delta\mu$-mixing.  The \sxO\  
depletion is a trivial 0.25\% (Figure~13).

In low-mass, low-metallicity stars, the $\delta\mu$-mixing 
brings the effective bottom of the convection zone to a 
point where \sxO\ and \ttNa\ are beginning to change. Because of 
this the change in these isotopes is dependent on the location 
that the mixing model finds for the base of the $\delta\mu$-mixing, 
where the mixing and burning of \tHe\ are in balance.  The 
mixing model developed here does not appear sufficient to explain 
an observable \sxO\ depletion or \ttNa\ enhancement, but this result 
warrants additional investigation. Because of the temperature 
sensitivity, a modest amount of (downwards) overshoot, or turbulent mixing, 
could change this result. Indeed, we note that our $\delta
\mu$-mixing mechanism makes it far easier for other mixing mechanisms to
have an effect. In the absence of $\delta \mu$-mixing there is a
radiative region of approximately $1\Rsun$ (or $0.02\Msun$ or 10
pressure scale-heights) between the hydrogen shell and the convective
envelope. Any proposed mixing mechanism must lift material form the
shell through this radiative region to make contact with the convective
envelope. However, with $\delta \mu$-mixing acting from just above the
shell, and linking to the convective envelope, the radiative region
between the shell and the bottom of the mixed region is now reduced to
approximately $0.005\Rsun$ (or $0.0004\Msun$ or less than one pressure scale
height). Thus other forms of mixing may be aided by the
operation of $\delta \mu$-mixing.

\section{Mass Loss after the Mixing does not Matter}

As noted in the introduction, the blue horizontal branches of 
globular clusters indicate that a substantial fraction of the 
envelope is lost prior to the helium core flash.  This was the 
origin of the excess \tHe\ production by these stars.  With this 
new mixing mechanism, once the $\mu$-inversion occurs, the \tHe\ 
abundance drops rapidly.  As the mixing depends on the square 
of the \tHe\ abundance, the process slows dramatically when 90\% 
of this isotope has been consumed.  Mass loss after this rapid 
drop does not cause contamination problems with excess \tHe, 
and leads to little change in the surface abundances or yields.  
To illustrate this, a 0.8$\Msun$ $Z$ = 0.0001 model was run with no mass 
loss, and then again with 0.2$\Msun$ of mass loss near the tip off the 
giant branch (and after the $\delta\mu$-mixing). The 
surface abundances show little change from the mass loss.  

In the absence of $\delta\mu$-mixing, the mass losing 
model ejects 52 times as much \tHe\ into the ISM as it took when it 
formed. This was the basis of the problem of reconciling Big 
Bang nucleosynthesis with yields from stellar evolution. With
 $\delta\mu$-mixing, the mass loss ejects 1/3rd of 
the \tHe\ into the ISM as it took when the star formed, and 
retains only 1/7th of the intake to be further processed or ejected. 

\section{Conclusions}

\par Our first conclusion is that $\delta\mu$-mixing is a significant and 
inevitable 
process in low-mass stars ascending the giant branch for the first time.  
Once it begins, the timescale is short (compared with the overall nuclear
timescale of $\sim 10^8\thn$yrs, and it maintains a nearly homogeneous 
composition down to base temperatures in the region of 16 to 25 million 
K, allowing nuclear processing.  The result is an observable 
change in the expected abundances of \tHe\  and the CNO isotopes.

\par This mixing mechanism is driven by the destruction of \tHe, and is 
self-limiting.  The lowest mass stars ($< 1.25\Msun$), that were expected to 
produce a problematic excess of \tHe, quickly destroy 90 to 95\% of that 
isotope. As a result, the \tHe\  returned to the ISM is within the limits 
posed by Hata et al (1995).  This mixing also modifies the \twC/\thC\ ratios.  
Instead of showing a significant mass dependence in that ratio, we find 
a metallicity dependence instead.  Shortly after this mixing begins, Pop I 
stars between 0.8 and 2.0$\Msun$ should all drop to a value near 
14.5. Extreme Pop II stars in this mass range should show ratios near 4. 
Likewise the nitrogen and oxygen isotope ratios are substantially affected 
(Figure~14).

\par As a first effort to develop a 1D model for this mixing process, we have 
tried to validate it with reference to the original 3D modeling, and basic 
physical arguments.  At the moment, variations due to mesh resolution and 
mixing speed suggest an uncertainty of $\pm 2$ for Pop I values of the 
\twC/\thC\ ratio (much less for Pop II, where it reaches its equilibrium
value of $\sim 4$).  The fractional destruction of 
\tHe\  seems less sensitive to such choices.  We have also held the initial 
isotope ratios fixed (\twC/\thC\ = 90,  \frN/\ffN\ = 270, \sxO/\svO\ = 2625, 
and \sxO/\etO\ = 490), and different values must be run to study chemical 
evolution.

Although our diffusive model of $\dm$-mixing is very tentative, we have
endeavored to show that the details are not very important. We
estimate mixing speeds between 0.2 and 2 cm/s in the burning region,
increasing to perhaps 1 m/sec in the radiative zone below the convective
envelope. We also showed that any rate
within one or two orders of magnitude of the rate we used is going to produce 
much the same result so far as mixing is concerned. It is rather
unsatisfying not to have a more accurate estimate of the speed of
mixing. However, for the problems discussed in this paper the speed is
not crucial, as long as it is fast relative to the nuclear timescale,
and certainly this is what we find, by some orders of magnitude. For
other applications of the same mechanism it may be that a more accurate
estimate of the speed is required. We note that appealing to current
theories of thermohaline mixing is unlikely to help, as these vary from
one author to the other by two orders of magnitude. Further, since the
mixing timescale is essentially the Kelvin-Helmholtz timescale, then a
direct numerical simulation is impossible at present.

We believe that our model 
accounts remarkably well for at least two results: the fact that $^3$He is
not as much enhanced in the ISM as earlier models suggested, and the fact that
the $^{12}$C/$^{13}$C is seen to decrease substantially further than was
found in earlier models. We expect in a later paper to look in more detail at 
oxygen isotopes as well as $^7$Li in red-giants.

\par Finally, this process does not appear to have significant impact on the 
\sxO\  depletion or \ttNa\  enhancement, but with the temperature sensitivity 
(and rate uncertainties)
of these rates, it comes close.  Any overshoot in the mixing at the bottom 
of the region will be important.  Also, the thickness of the stable region 
that surrounds the hydrogen burning shell is reduced from over a solar 
radius to a few hundredths of a solar radius.  This may enable other 
mechanisms (rotation, magnetic fields, ...) to create variation in 
the observed abundances. Alternatively, the homogeneity seen in clusters like 
M67 (Gilroy \& Brown 1991) might be used to limit models for such mixing.


\section{Acknowledgments} 

This study has been carried out under the auspices of the U.S. Department of  
Energy, National Nuclear Security Administration, by the University of  
California, Lawrence Livermore National Laboratory, under contract  No. 
W-7405-Eng-48. JCL was partially supported by the Australian Research Council. 
We are indebted to R. Palasek for  assistance with the code..

\begin{figure}
\epsscale{.80}
\plotone{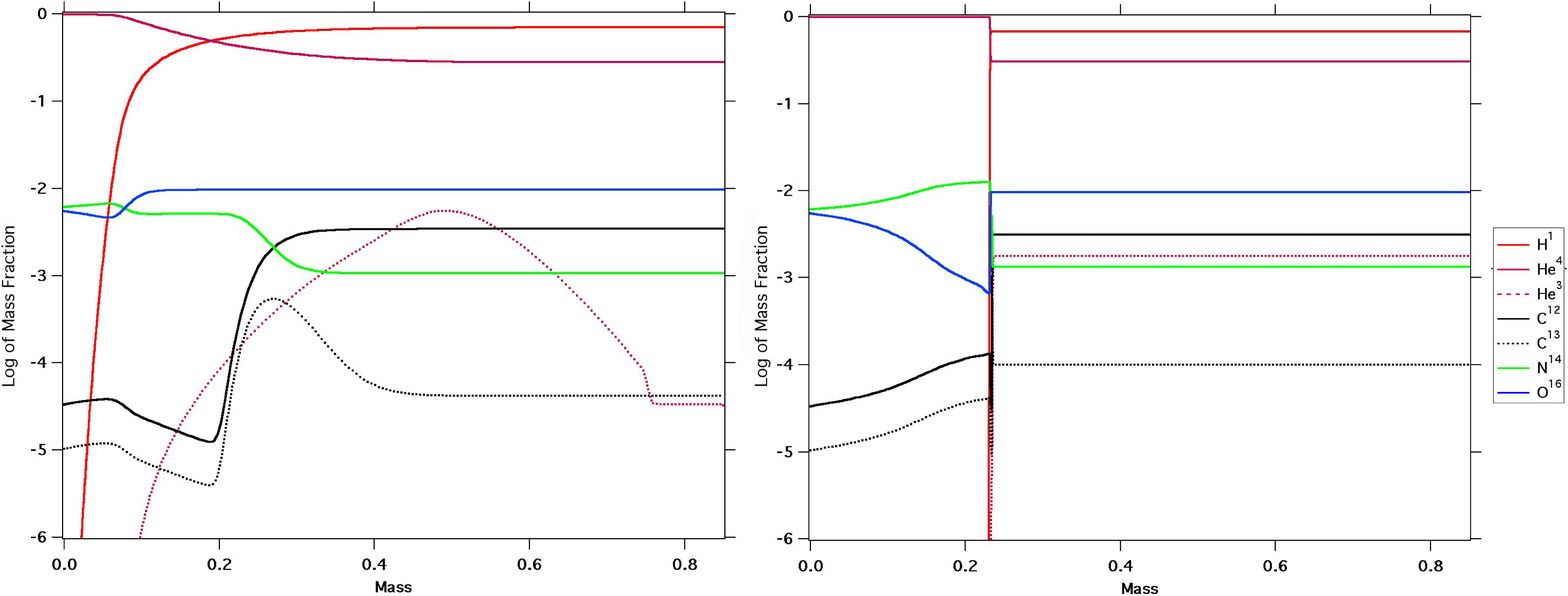}
\caption{Composition structure of a $0.85\Msun$ Pop I model with $X=0.7$,
$Z=0.02$. Left panel -- near the end of its main-sequence life. 
Right panel -- near the deepest penetration of 
the surface convective region on the giant branch. \oH\ -- vermilion;
\fHe\ -- magenta; \twC\ -- black; \sxO\ -- blue; \frN\ -- green; \tHe\ 
-- red dots; \thC\ -- black dots.}
\end{figure}
\clearpage
\begin{figure}
\epsscale{.80}
\plotone{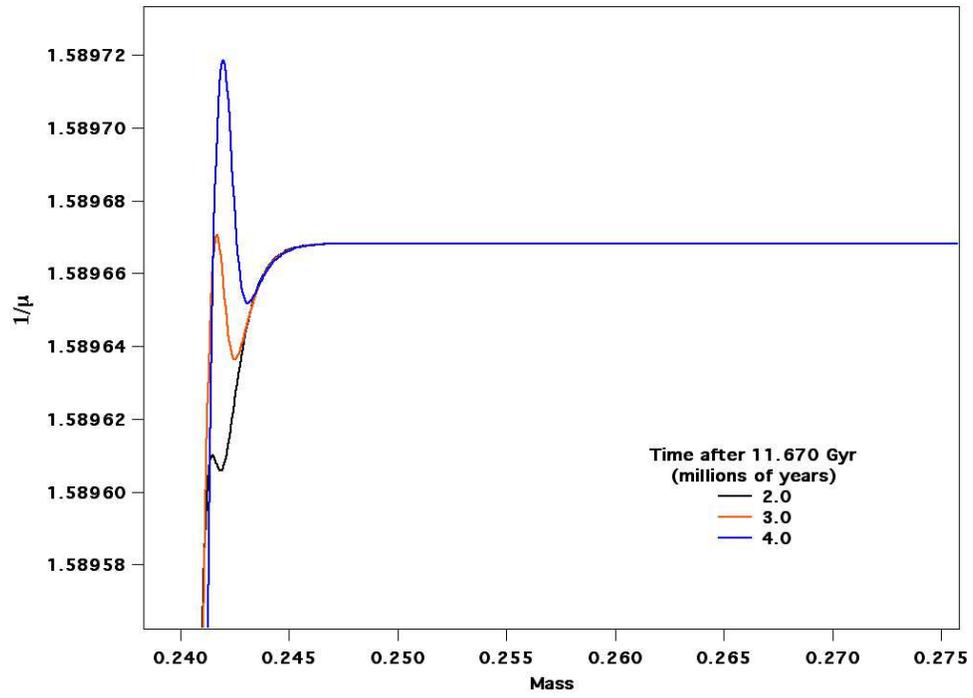}
\caption{The development of a molecular-weight inversion is shown 
(versus mass) for a $1.0\Msun$ Pop I model with $X=0.7$, $Z=0.02$.  $1/\mu$ 
curves are shown for million-year increments beginning at 11.67~Gyr.}
\end{figure}
\clearpage

\begin{figure}
\epsscale{.80}
\plotone{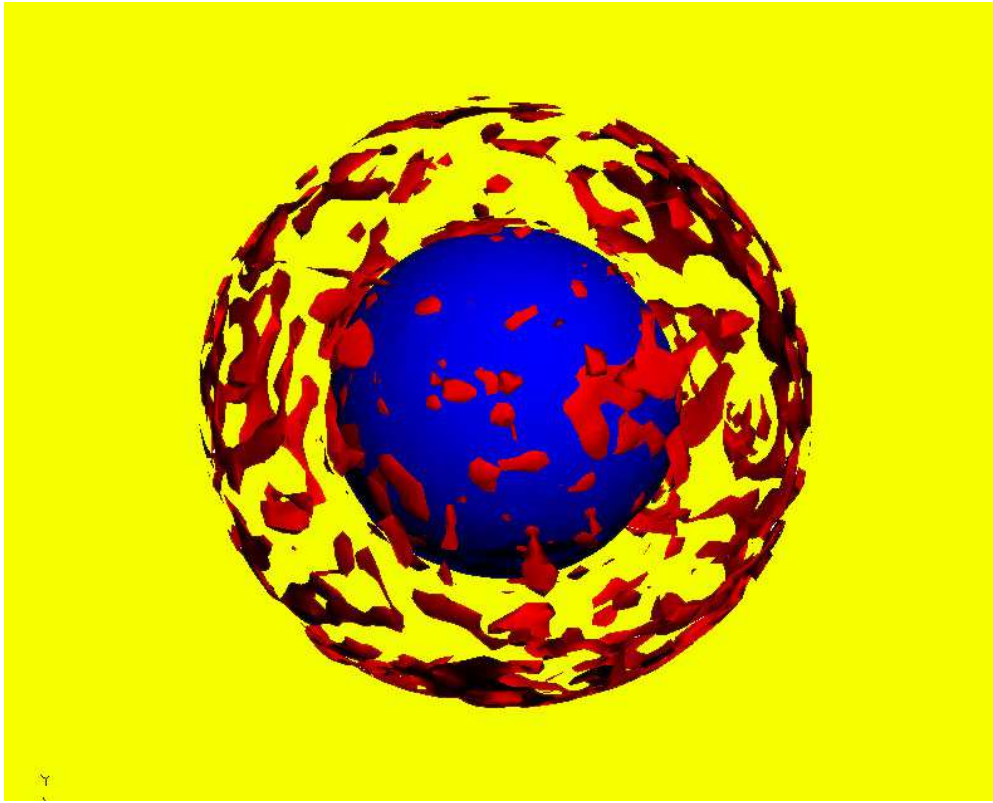}
\caption{The location of the hydrogen burning shell is shown by a blue 
contour at a fixed mass fraction of $^{13}$N. Shown in red are rising clouds 
in which the hydrogen abundance is marginally higher than in the 
surrounding material.}
\end{figure}
\clearpage

\begin{figure}
\epsscale{.80}
\plotone{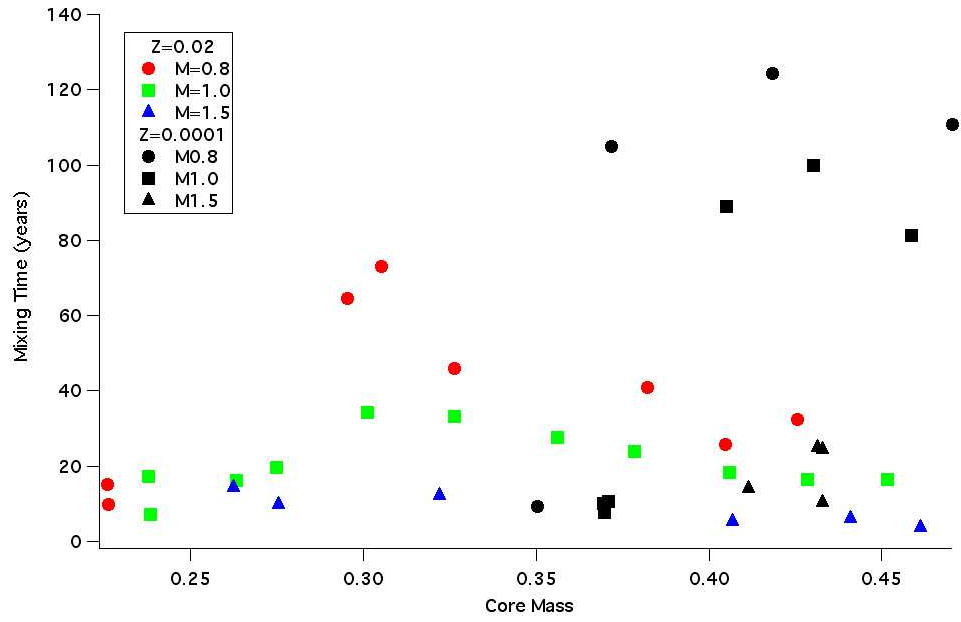}
\caption{Mixing times along the giant branch, for Pop I and Pop II
models with a range of masses.}
\end{figure}
\clearpage

\begin{figure}
\epsscale{.80}
\plotone{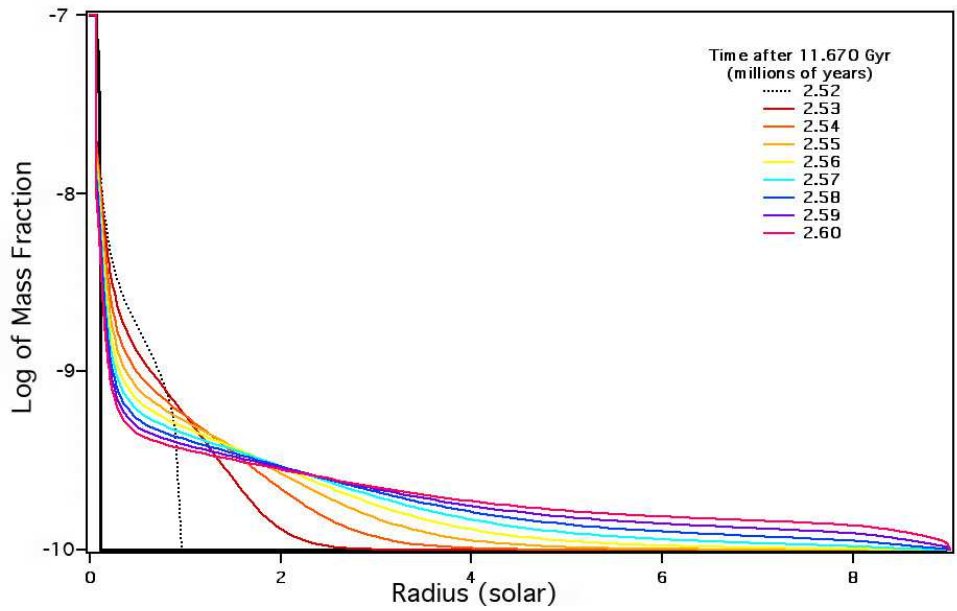}
\caption{The mass fraction profile created by the proposed diffusion 
coefficient in a $1\Msun$ star ($Z=0.02$), starting from a pure step-function
profile. Each curve is separated by 10,000 
years.  For the value of $\finv$\  selected here (100), the speed at which the 
material is carried outward corresponds to about 0.3 cm/s.}
\end{figure}
\clearpage 

\begin{figure}
\epsscale{.80}
\plotone{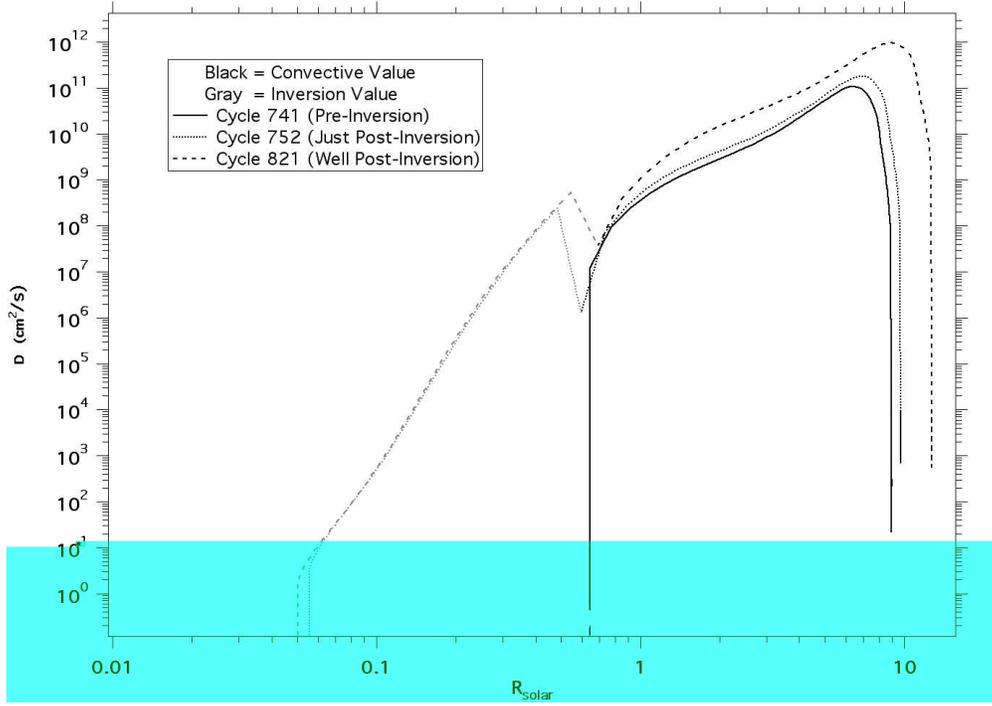}
\caption{The diffusion co-efficient that we use 
(in cm$^2$/s), (a) solid line, cycle 741, before our mechanism sets in, 
so that there is only the normal surface convection zone ($D$ determined
by equation 14); 
(b) dotted, cycle 752, just
after $\delta\mu$-mixing begins, $D$ from equation (15) except for the
convective region, which uses equation (14); (c) dashed, cycle 821, well
after the mixing is established.
Radius is in solar units.}
\end{figure}
\clearpage

\begin{figure}
\epsscale{.80}
\plotone{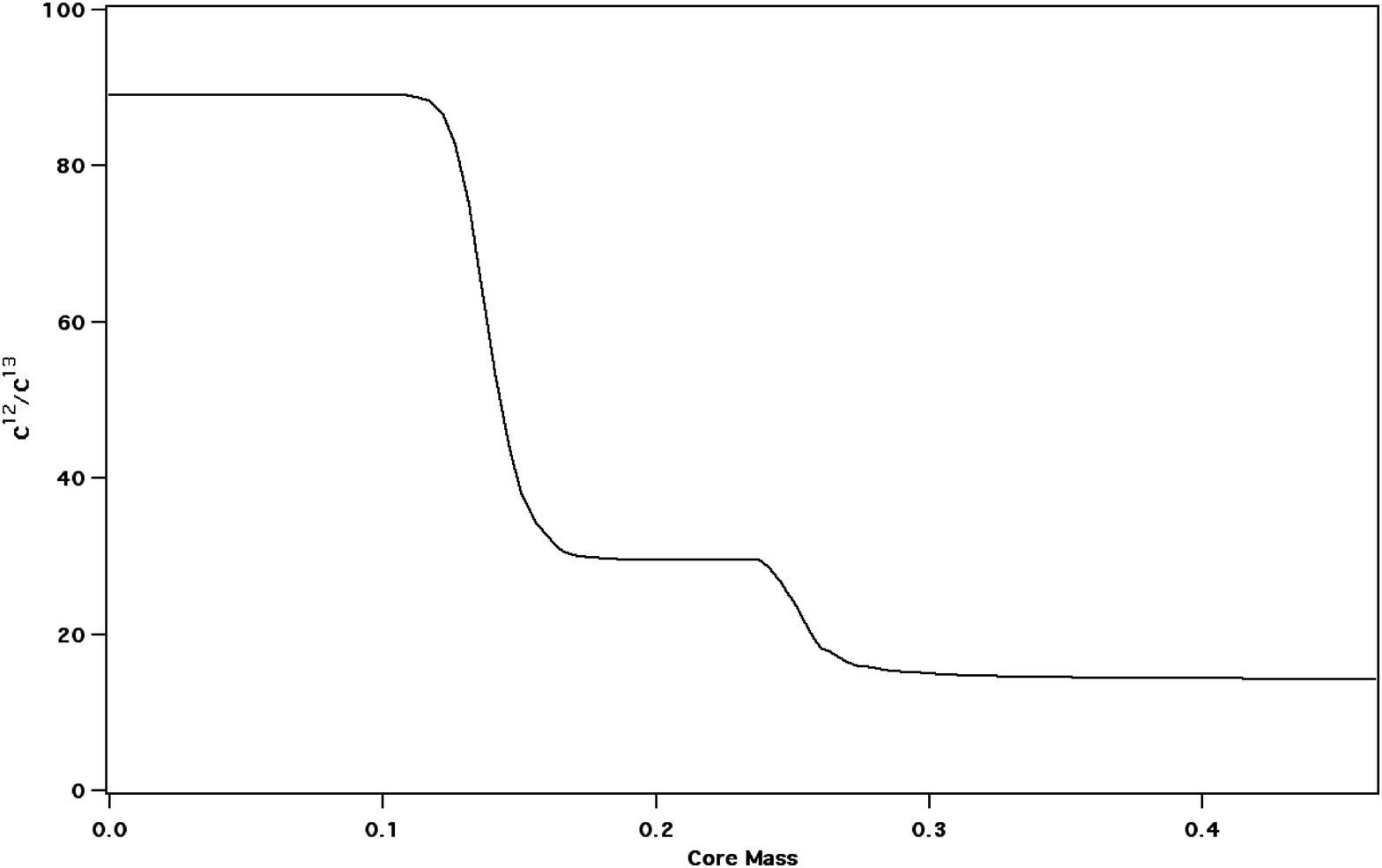}
\caption{The \twC/\thC\  ratio versus core mass in a $1\Msun$ model.}
\end{figure}
\clearpage

\begin{figure}
\epsscale{.80}
\plotone{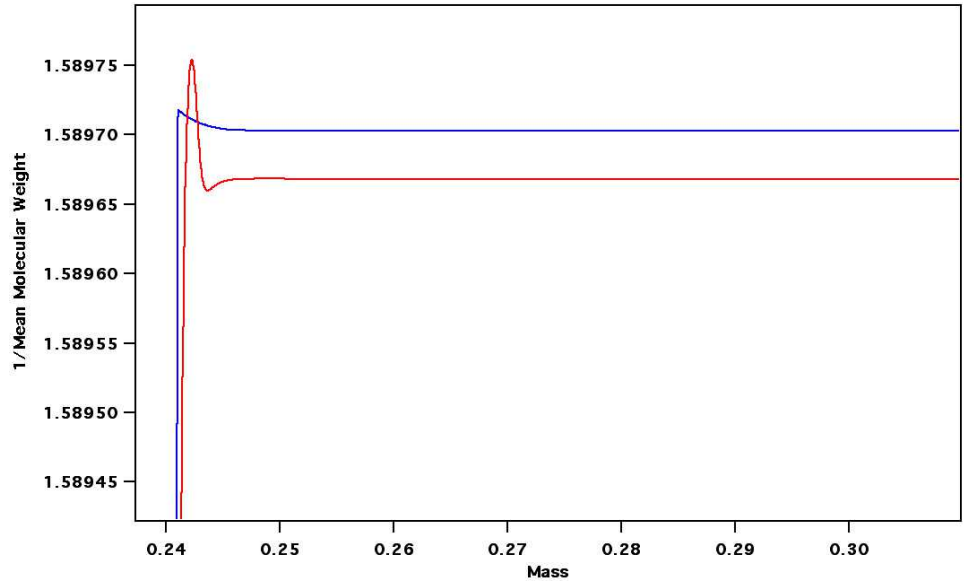}
\caption{Approximately 3 million years after the hydrogen burning shell 
approaches the homogeneous region, the $1/\mu$ profiles for a $1\Msun$ model 
with no mixing (red) and a $\delta\mu$-mixing model (blue).}
\end{figure}
\clearpage

\begin{figure}
\epsscale{.80}
\plotone{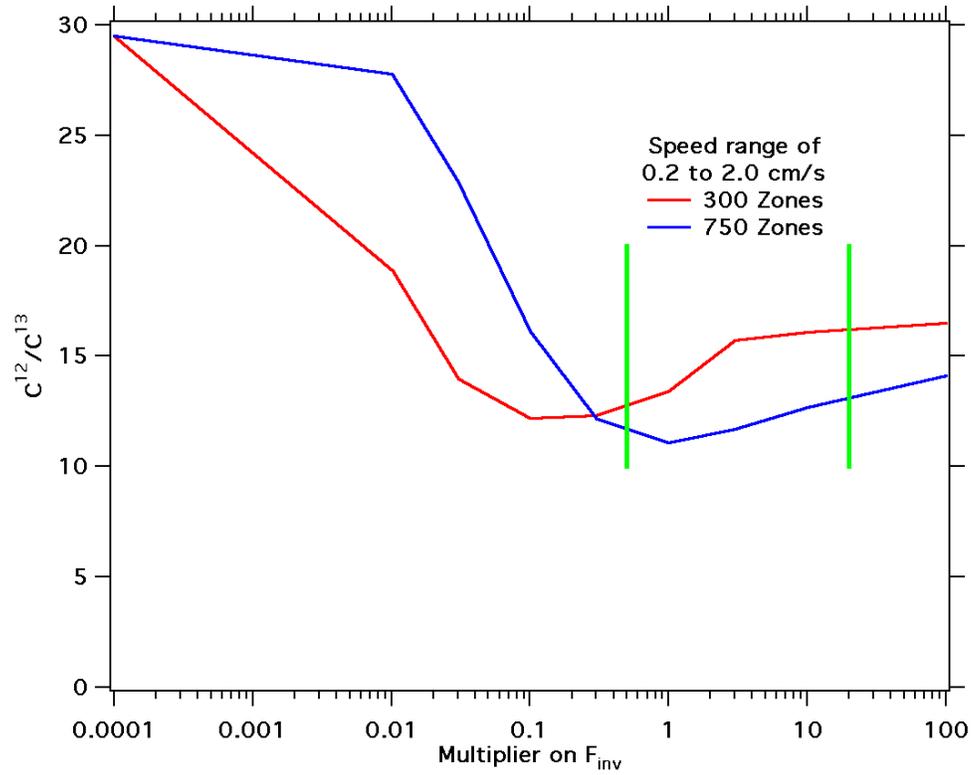}
\caption{Surface carbon isotope ratios as a function of $\finv$\ 
and the number of zones used in the 1D model. The region corresponding
to the estimated speed range of 0.2 to 2 cm/sec is shown by the
vertical green lines. The x-axis is the multiplier used on
the standard $\finv =100$.}
\end{figure}
\clearpage

\begin{figure}
\epsscale{.80}
\plotone{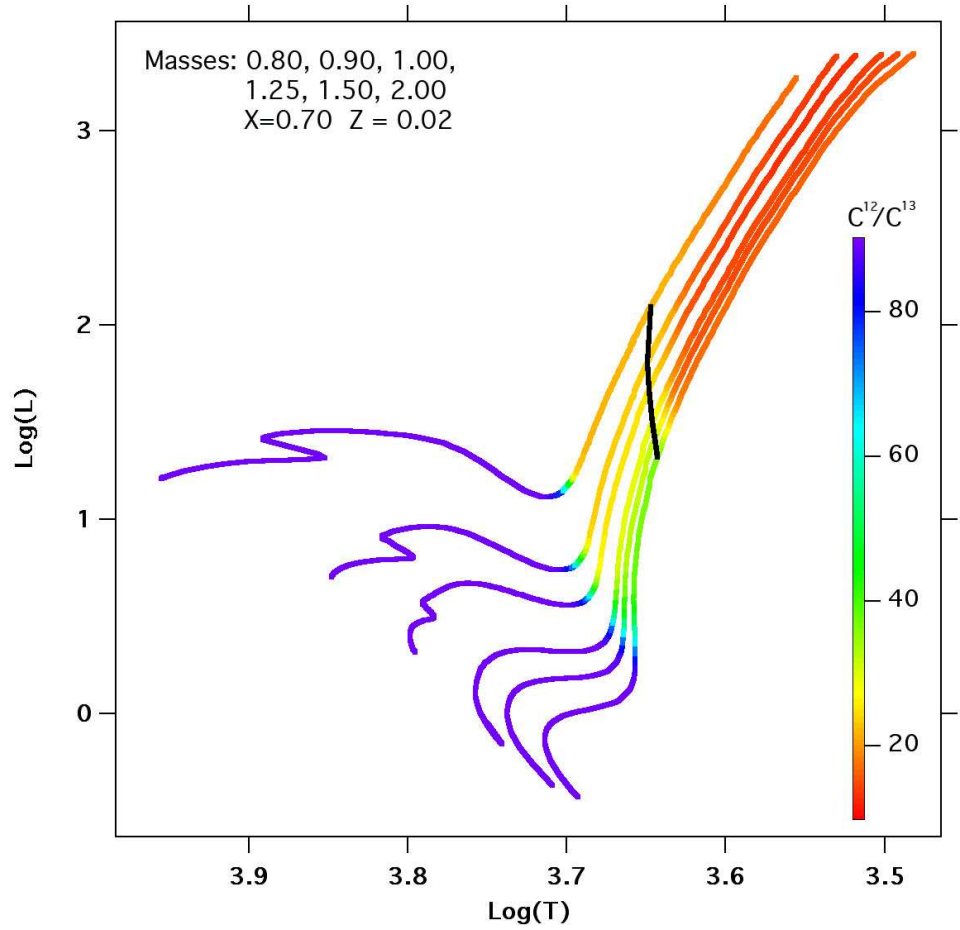}
\caption{\twC/\thC\  ratio for various masses using our standard 
$\finv =100$.  The nearly vertical dark line marks the onset of the $\delta \mu$-mixing process.}
\end{figure}
\clearpage
\begin{figure}
\epsscale{.80}
\plotone{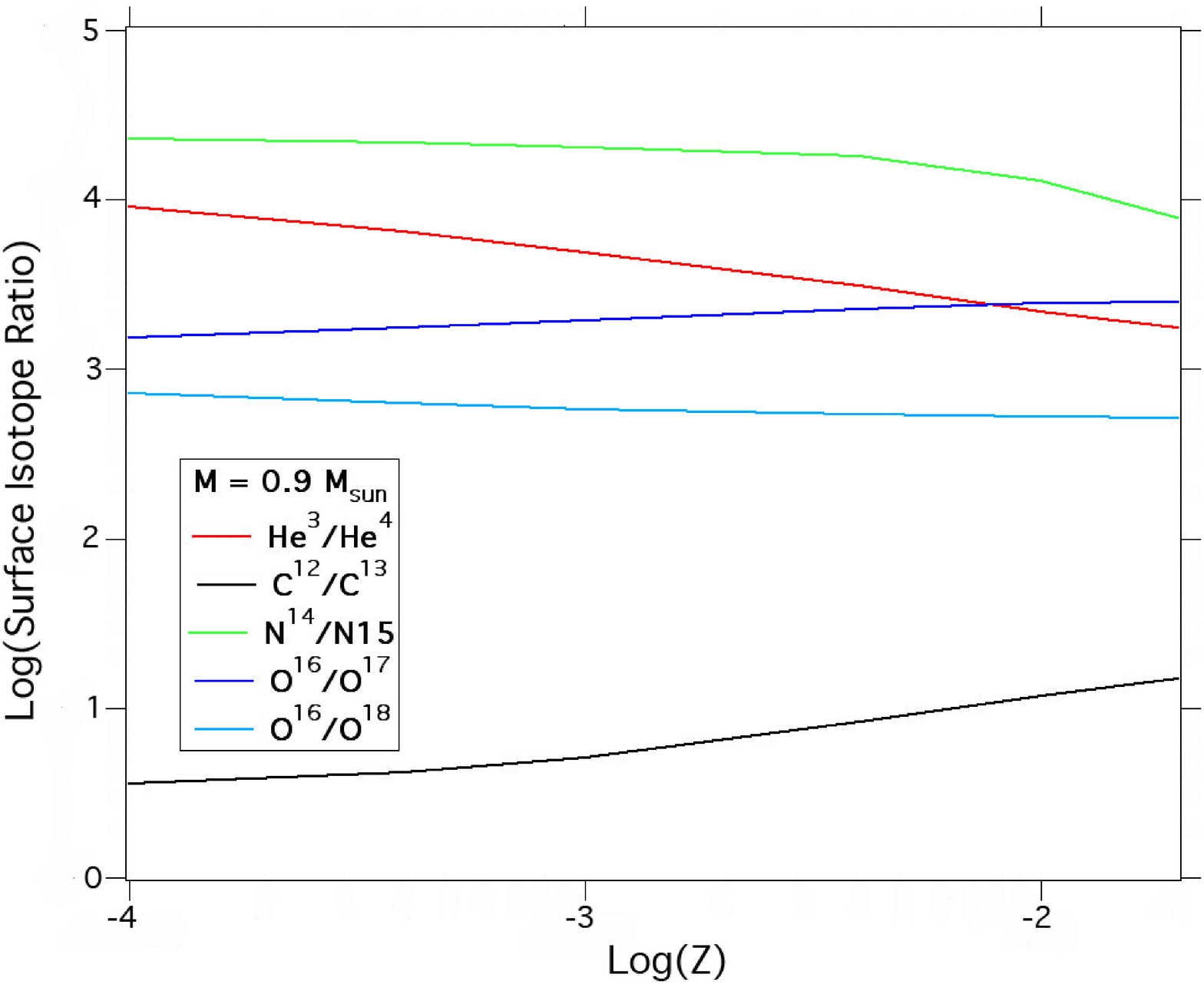}
\caption{Post-mixing, surface isotope ratios in a $0.9\Msun$ model with 
a large range of initial abundances.}
\end{figure}
\clearpage
\begin{figure}
\epsscale{.80}
\plotone{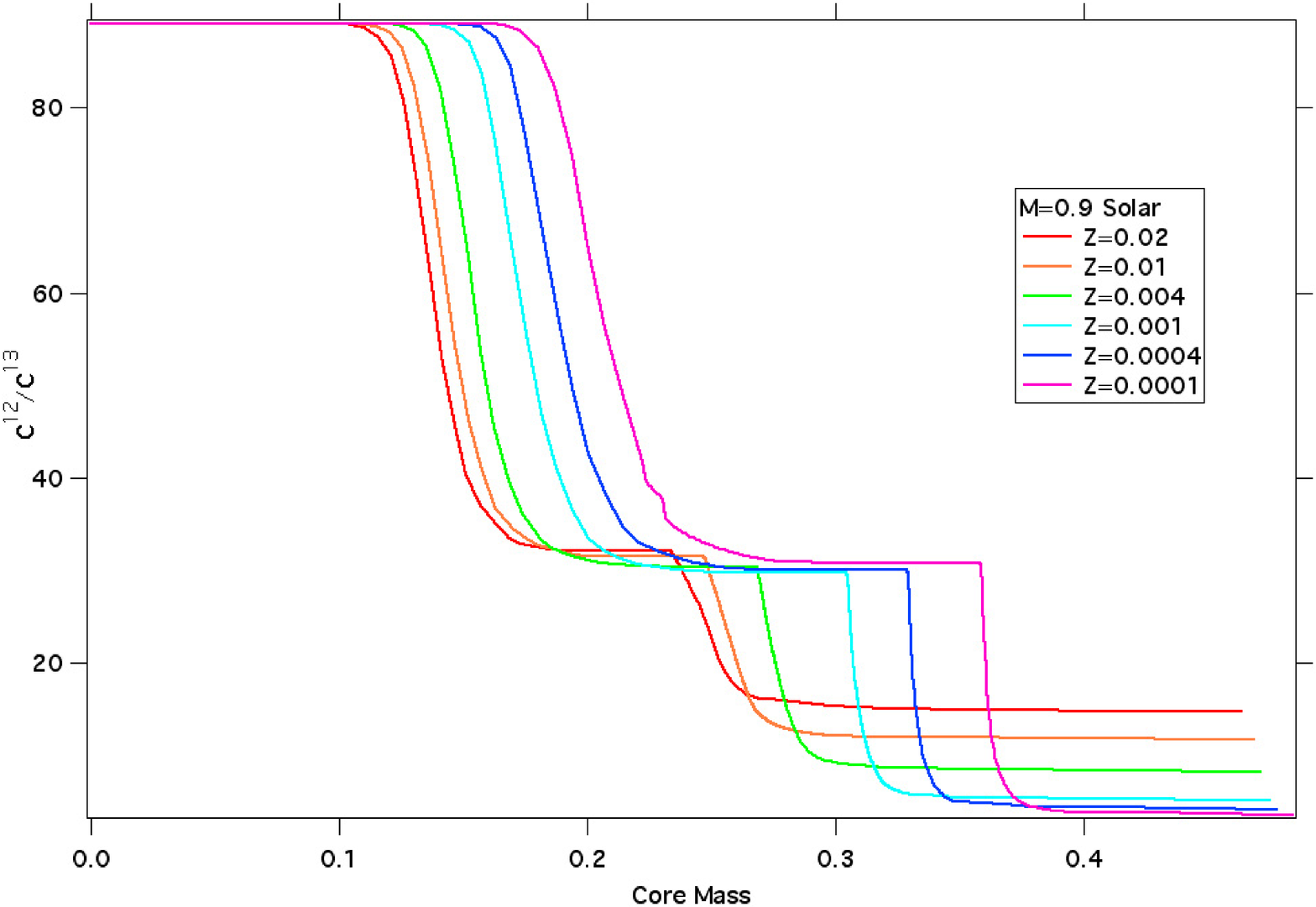}
\caption{The \twC/\thC\  ratio verses core mass for a range of 
metallicities.}
\end{figure}
\clearpage
\begin{figure}
\epsscale{.80}
\plotone{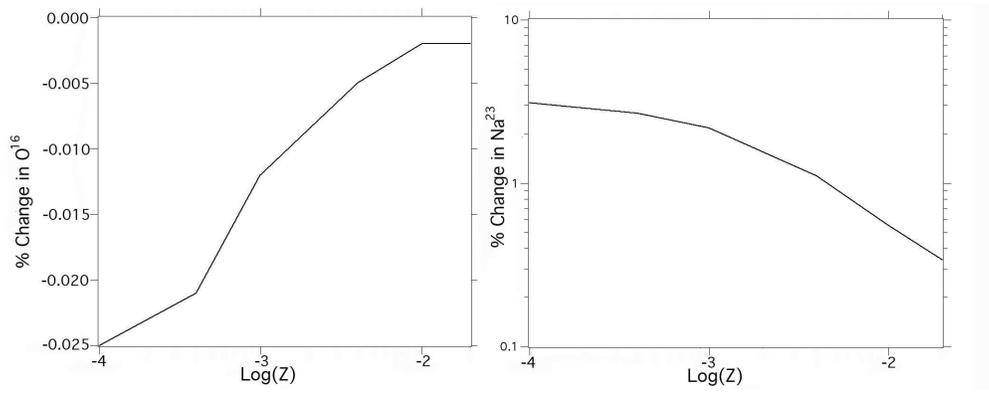}
\caption{Change in the surface abundance of \sxO\ and \ttNa\ in 
a $0.9\Msun$ model evolved with various metallicities.}
\end{figure}
\clearpage
\begin{figure}
\epsscale{.80}
\plotone{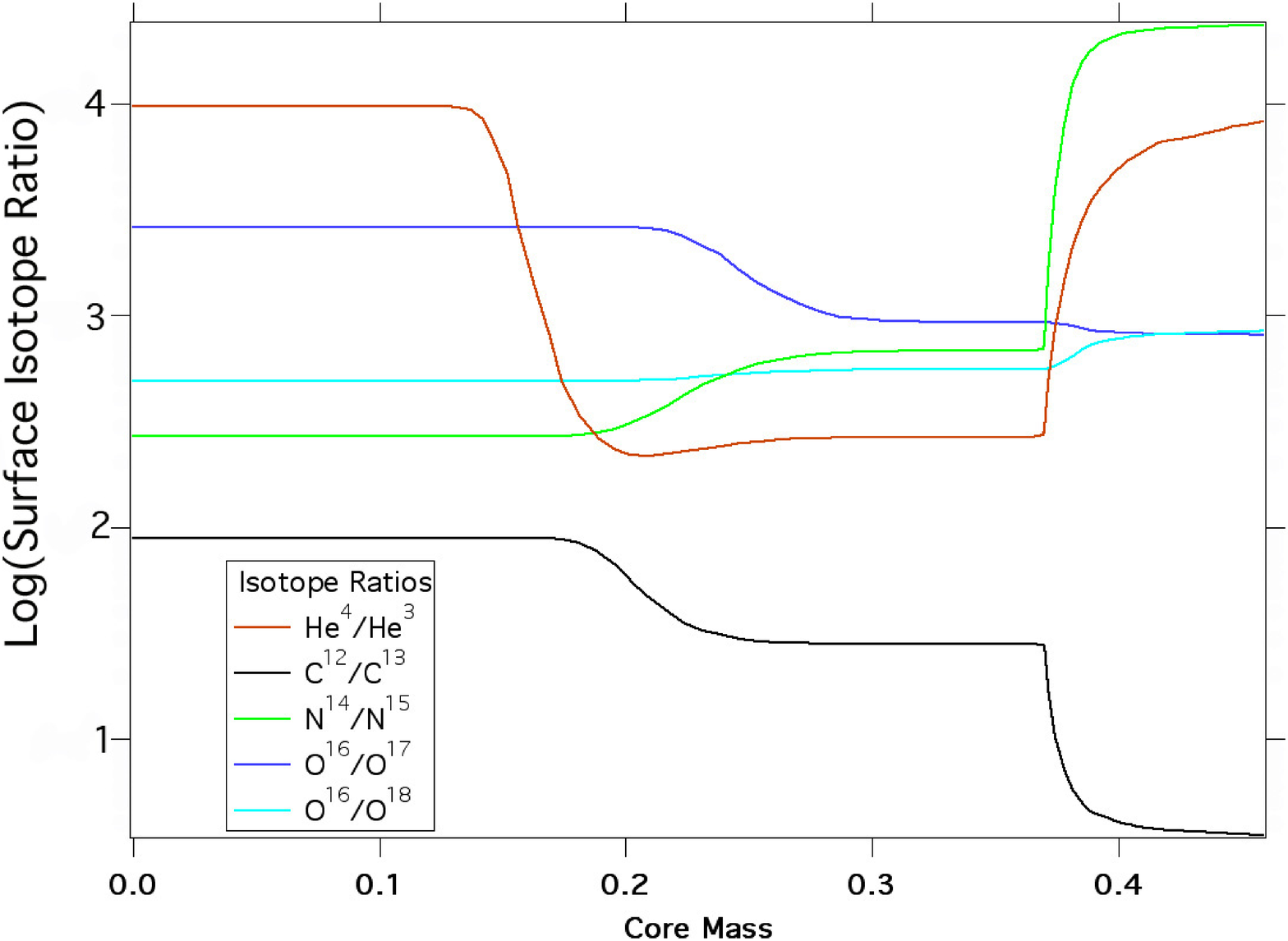}
\caption{CNO isotope ratios in a $1\Msun$ ($Z$ = 0.0001) model are 
plotted against core mass.}
\end{figure}
\clearpage

\begin{deluxetable}{ccccc}
\tablecaption{Effect of Varying $\finv$ and Number of Zones }
\tablewidth{0pt}
\startdata
\tableline\\
      &\w\w 300 &Zones\w\w    &\w\w 750 &Zones\w\w\\
$\finv$&\twC/\thC&\tHe\ Destroyed&\twC/\thC&\tHe\ Destroyed\\
\tableline\\
0     &   29.5  &     0\%     &  29.1   &     0\%   \\
1          &   18.9  &  59.0\%     &  27.8   &  35.8\%   \\
3          &   14.0  &  75.8\%     &  22.9   &  46.7\%   \\
10         &   12.2  &  87.3\%     &  16.1   &  64.9\%   \\
30         &   12.3  &  91.3\%     &  12.2   &  80.6\%   \\ 
100        &   13.4  &  92.6\%     &  11.1   &  89.1\%   \\
300        &   15.7  &  92.4\%     &  11.7   &  91.9\%   \\
10$^3$     &   16.1  &  92.8\%     &  12.7   &  93.1\%   \\
10$^4$     &   16.5  &  93.9\%     &  14.1   &  93.8\%   \\
\enddata
\end{deluxetable}
\clearpage

\begin{deluxetable}{cccccc}
\tablecaption{Effect of varying mass and $\finv$}
\tablewidth{0pt}
\startdata
\tableline\\
         &           &\x $\finv=10^2$&\x\x\x $\finv=10^2$     &\x\x $\finv=10^4$&\x\x\x $\finv=10^4$\\
Mass	 & \x FDU & \x\twC/\thC&\x\tHe\ Destroyed &\x\x\twC/\thC&\x\tHe\ Destroyed\\
\tableline\\
0.80	 & \x  36.9  & \x 15.9 &\x\x\x 96.4\%     &\x\x 19.5 &\x\x\x 96.7\%    \\
0.85	 & \x  34.0  & \x 15.3 &\x\x\x 95.7\%     &\x\x 18.5 &\x\x\x 96.0\%    \\
0.90	 & \x  32.2  & \x 14.5 &\x\x\x 94.8\%     &\x\x 17.6 &\x\x\x 95.4\%    \\
1.00	 & \x  29.5  & \x 13.4 &\x\x\x 92.6\%     &\x\x 16.5 &\x\x\x 93.5\%    \\
1.25	 & \x  25.6  & \x 13.0 &\x\x\x 85.7\%     &\x\x 14.9 &\x\x\x 89.1\%    \\
1.50	 & \x  23.6  & \x 13.7 &\x\x\x 74.9\%     &\x\x 14.4 &\x\x\x 82.3\%    \\
2.00	 & \x  22.3  & \x 17.0 &\x\x\x 45.1\%     &\x\x 14.9 &\x\x\x 63.6\%    \\
\enddata
\end{deluxetable}
\clearpage

\begin{deluxetable}{ccccccc}
\tablecaption{\twC/\thC\ ratios}
\tablewidth{0pt}
\startdata
\tableline\\
&\x X=0.70,   & Z=0.02   &\x X=0.738,   & Z=0.001 &\x X=0.74,  & Z=0.0004  \\
Mass\ &\x FDU     &\x Final     &\x FDU &\x Final &\x FDU &\x Final \\
0.80&\x 36.9&\x 15.9&\x 34.1&\x 5.3&\x 35.0&\x	4.2\\
0.85&\x 34.0&\x 15.3&\x 31.5&\x 5.0&\x 31.8&\x	4.0\\
0.90&\x 32.2&\x 14.5&\x 29.6&\x 4.9&\x 30.0&\x	4.0\\
1.00&\x 29.5&\x 13.4&\x 27.3&\x 4.9&\x 27.4&\x	4.0\\
1.25&\x 25.6&\x 13.0&\x 24.3&\x 5.0&\x 24.3&\x	4.1\\
1.50&\x 23.6&\x 13.7&\x 24.3&\x 5.2&\x 22.7&\x	4.6\\
2.00&\x 22.3&\x 17.0&\x 21.2&\x 14.2&\x 21.0&\x 21.0\\
\enddata
\end{deluxetable}

\clearpage
\begin{deluxetable}{cccccc}
\tablecaption{Surface Isotope Ratios at the Top of the Giant Branch}
\tablewidth{0pt}
\tablehead{
\colhead{Z} & \colhead{\tHe/\fHe} & \colhead{\twC/\thC} & \colhead{\frN/\ffN} & \colhead{\sxO/\svO} &
\colhead{\sxO/\etO}
}
\startdata
0.02    \w&1.8$\by 10^3$\w&\z14.5     \w&0.8$\by 10^4$\w&2.5$\by 10^3$\w&\z513.\\
0.01	\w&2.3$\by 10^3$\w&\z11.3     \w&1.4$\by 10^4$\w&2.4$\by 10^3$\w&\z521.\\
0.004	\w&3.2$\by 10^3$\w&\z8.0      \w&1.8$\by 10^4$\w&2.3$\by 10^3$\w&\z537.\\
0.001	\w&5.2$\by 10^3$\w&\z4.9      \w&2.1$\by 10^4$\w&1.9$\by 10^3$\w&\z584.\\
0.0004	\w&6.8$\by 10^3$\w&\z4.0      \w&2.2$\by 10^4$\w&1.7$\by 10^3$\w&\z640.\\
0.0001	\w&9.1$\by 10^3$\w&\z3.5      \w&2.3$\by 10^4$\w&1.5$\by 10^3$\w&\z752.\\
\enddata
\end{deluxetable}

\clearpage
\begin{deluxetable}{ccccccc}
\tablecaption{\tHe\ production in a $0.9\Msun$ model.}
\tablewidth{0pt}
\tablehead{
\colhead{} & \colhead{} & \colhead{} & \colhead{} & \colhead{} &
\colhead{}
}
\startdata
$Z$	\w&0.02  \w&0.01  \w&0.004 \w&0.001 \w&0.0004\w&0.0001\\
Peak	\w&61.1  \w&56.5  \w&48.7  \w&40.0  \w&38.7  \w&38.1  \\
Final	\w&3.1   \w&2.3   \w&1.6   \w&1.0   \w&0.7   \w&0.5   \\
Change	\w&94.8\%\w&95.8\%\w&96.7\%\w&97.6\%\w&98.1\%\w&98.6\%\\
\enddata
\end{deluxetable}

\clearpage
\begin{deluxetable}{ccccccc}
\tablecaption{\tHe/\tHe\ (original)}
\tablewidth{0pt}
\startdata
\tableline\\
          &\x X=0.70,& Z=0.02\w\w &\x X=0.738, & Z=0.001\w\w &\x X=0.74,  & Z=0.0004\w\w  \\
$M$     \w&Peak	\w&Mixed\w&Peak	        \w&Mixed	\w&Peak	\w&Mixed\\
0.80	\w&76.6	\w&2.7	\w&54.7 	\w&0.86 	\w&53.0	\w&0.62\\
0.85	\w&68.2	\w&2.9	\w&46.5 	\w&0.90 	\w&45.1	\w&0.66\\
0.90	\w&61.1	\w&3.1	\w&40.0	        \w&0.95 	\w&38.7	\w&0.73\\
1.00	\w&49.7	\w&3.7	\w&32.9 	\w&1.12 	\w&30.0	\w&0.83\\
1.25	\w&31.7	\w&4.5	\w&21.8 	\w&1.60 	\w&19.9	\w&1.21\\
1.50	\w&21.8	\w&5.5	\w&15.7 	\w&2.08 	\w&14.4	\w&1.91\\
2.00	\w&12.8	\w&7.0	\w&9.5  	\w&6.70 	\w&8.7 	\w&8.39\\
\enddata
\end{deluxetable}

\clearpage

\begin{deluxetable}{ccc}
\tablecaption{Enhancements with and without Mass Loss}
\tablewidth{0pt}
\tablehead{
\colhead{Isotope Ratios} & \colhead{No Loss} & \colhead{Loss}
}
\startdata
\tHe/\fHe	\w&9065.5	  \w&9466.7\\
\twC/\thC	\w&3.5	  \w&3.5\\
\frN/\ffN	\w&22564.1  \w&22330.\\
\sxO/\svO	\w&1913.4   \w&1920.8 \\
\sxO/\etO	\w&702.6    \w&677.8 \\
$\delta$(\sxO)	\w& -0.01\% \w&-0.01\%\\
$\delta$(\ttNa)	\w&1.2\%    \w&1.3\%\\
\enddata
\end{deluxetable}

\clearpage
\end{document}